\documentstyle[11pt]{article}
\begin{document}\bibliographystyle{plain}
\begin{titlepage}\renewcommand{\thefootnote}{\fnsymbol{footnote}}
\hfill\begin{tabular}{l}HEPHY-PUB 728/00\\UWThPh-2000-8\\hep-ph/0002139\\
February 2000\end{tabular}\\[3cm]\Large\begin{center}{\bf NUMERICAL SOLUTION
OF THE\\SPINLESS SALPETER EQUATION\\BY A SEMIANALYTICAL MATRIX METHOD\\(A
Mathematica 4.0 Routine)}\\\vspace{2cm}\large{\bf Wolfgang
LUCHA\footnote[1]{\normalsize\ {\em E-mail\/}:
wolfgang.lucha@oeaw.ac.at}}\\[.5cm]Institut f\"ur
Hochenergiephysik,\\\"Osterreichische Akademie der
Wissenschaften,\\Nikolsdorfergasse 18, A-1050 Wien, Austria\\[2cm]{\bf Franz
F. SCH\"OBERL\footnote[2]{\normalsize\ {\em E-mail\/}:
franz.schoeberl@univie.ac.at}}\\[.5cm]Institut f\"ur Theoretische
Physik,\\Universit\"at Wien,\\Boltzmanngasse 5, A-1090 Wien,
Austria\vfill{\normalsize\bf Abstract}\end{center}\normalsize

In quantum theory, the ``spinless Salpeter equation,'' the relativistic
generalization of~the nonrelativistic Schr\"odinger equation, is used to
describe both bound states of scalar particles and the spin-averaged spectra
of bound states of fermions. A numerical procedure solves the spinless
Salpeter equation by approximating this eigenvalue equation by a matrix
eigenvalue problem with {\em explicitly known matrices\/}.

\vspace{6ex}

\renewcommand{\thefootnote}{\arabic{footnote}}\end{titlepage}

\normalsize

\section{Introduction}The appropriate framework for the description of bound
states within a relativistic quantum field theory is, beyond doubt, the
famous Bethe--Salpeter equation \cite{BSE}. Its actual application, however,
faces several well-known problems of both conceptual as well as practical
nature. For this reason, one usually considers some (three-dimensional)
reduction of this formalism. In particular, assuming a static interaction
between the bound-state constituents, neglecting all spin degrees of freedom,
and considering exclusively positive-energy solutions, one ends~up with what
is called the ``spinless Salpeter equation.'' For some brief accounts of the
reduction of the Bethe--Salpeter equation to the spinless Salpeter equation,
see, for instance, Refs.~\cite{Lucha92IJMPA,Lucha98:T}. This eigenvalue
equation generalizes the Schr\"odinger equation of the nonrelativistic
quantum theory towards relativistic kinematics. It thus accomplishes the
description of bound states~of spinless constituents (scalar bosons) as well
as of the spin-averaged spectra of bound states~of fermions, like, e.~g., in
elementary particle physics the description of hadrons as bound states of
quarks \cite{Lucha91:PRep} within the (intuitive) framework of potential
models for the strong interactions.

Here we present a very efficient method, based on Ref.~\cite{Lucha92NAM}, for
solving the spinless~Salpeter equation in configuration space, where the
interaction potentials are usually formulated. The efficiency of this method
stems from the fact that it does not require numerical integrations; rather,
the aim of this method is to employ analytical results whereever these are
available.

The outline of this paper is as follows. In Sect.~\ref{Sec:SSEIC}, the
configuration-space representation~of the spinless Salpeter equation is
briefly recalled. The resulting integro-differential equation is then
approximated, in Sect.~\ref{Sec:CEMEP}, by a (finite-dimensional) matrix
eigenvalue problem. The~only nontrivial ingredient of this matrix equation,
viz., the piece emerging from the kinetic~energy, is calculated in
Sect.~\ref{Sec:KEM}. This procedure for the determination of both energy
eigenvalues and corresponding wave functions of bound states described by a
spinless Salpeter equation~with an, in principle, arbitrary interaction
potential is eventually condensed, in Sect.~\ref{Sec:MNBS}, to a simple
Mathematica routine. (This routine may be obtained, of course, by contacting
the authors.)

\section{The Spinless Salpeter Equation in Configuration
Space}\label{Sec:SSEIC}For notational and conceptual simplicity, we consider
only the case of equal masses $m$ of~the bound-state constituents; the
generalization to the case of unequal masses is straightforward.

The semirelativistic Hamiltonian $H$ governing the dynamics of two
relativistically moving spinless particles of equal masses $m$, which
interact via some arbitrary coordinate-dependent static potential $V=V({\bf
x})$ reads, in the center-of-momentum frame of these two particles,
\begin{equation}H=2\,T+V\ ,\label{Eq:SRH}\end{equation}where $T$ denotes the
``square-root'' operator of the relativistic expression for the free
(kinetic) energy of some particle of mass $m$ and momentum ${\bf
p}$,\begin{equation}T=T({\bf p})\equiv\sqrt{{\bf p}^2+m^2}\ .\label{Eq:KEO}
\end{equation}The spinless Salpeter equation is nothing else but the
eigenvalue equation for the operator~$H$,
\begin{equation}H|\chi_k\rangle=E_k|\chi_k\rangle\ ,\quad k=0,1,2,\dots\
,\label{Eq:SSE}\end{equation}for the complete set of Hilbert-space
eigenvectors $|\chi_k\rangle$ with corresponding energy
eigenvalues$$E_k\equiv
\frac{\langle\chi_k|H|\chi_k\rangle}{\langle\chi_k|\chi_k\rangle}\ ,\quad
k=0,1,2,\dots\ .$$However, in contrast to, e.~g., the (nonrelativistic)
Schr\"odinger equation, the semirelativistic Hamiltonian $H$ is a nonlocal
operator, i.~e., either the relativistic kinetic-energy operator~$T$ in
configuration space or, in general, the interaction-potential operator $V$ in
momentum~space is nonlocal.

\newpage

In configuration space, the action of the kinetic-energy operator
(\ref{Eq:KEO}) on some element $\psi$ of $L_2(R^3)$, the Hilbert space of
square-integrable functions on the three-dimensional Euclidian space $R^3$,
is defined by$$(T\psi)({\bf x})=\frac{1}{(2\pi)^3}\int{\rm d}^3p\int{\rm
d}^3y\,\sqrt{{\bf p}^2+m^2}\,\exp[{\rm i}\,{\bf p}\cdot({\bf x}-{\bf
y})]\,\psi({\bf y})\ .$$With this definition, an integral representation of
the spinless Salpeter equation can be~found \cite{Nickisch84,Lucha92NAM}. We
focus our interest, for the moment, to spherically symmetric interaction
potentials $V({\bf x})=V(r),$ depending only on the radial coordinate
$r\equiv|{\bf x}|.$ For a generic eigenfunction~$\chi$ of $H$ describing a
state of definite orbital angular momentum $\ell,$ we introduce a reduced
radial wave function $w(r)$ by factorizing $\chi({\bf x})$ according to
$$\chi({\bf x})=r^\ell\,w(r)\,{\cal Y}_{\ell m}(\Omega)\ ,$$ where ${\cal
Y}_{\ell m}(\Omega)$ denote the (orthonormalized) spherical harmonics of
angular momentum $\ell$~and its projection $m$, depending on the solid angle
$\Omega$. For a nonvanishing mass of the bound-state constituents, that is,
$m>0,$ upon rescaling the radial variable $r$ according to $x:=m\,r$~and
introducing the scaled counterpart $\tilde w(x)$ of the reduced radial wave
function $w(r)$ by defining $w(r)=:m^{\ell+1}\,\tilde w(x)$, any dependence
of the eigenvalue equation (\ref{Eq:SSE}) on the mass $m$ of the two
bound-state constituents may be absorbed into a scaled (and therefore
dimensionless)~energy eigenvalue $\tilde E$ and interaction potential $\tilde
V(x)$:$$\tilde E:=\frac{E}{m}\ ,\quad \tilde V(x):=\frac{V(r)}{m}\ .$$The
spinless Salpeter equation (\ref{Eq:SSE}) is then equivalent to the
integro-differential
equation~\cite{Nickisch84,Lucha92NAM}\begin{equation}\left[\tilde E-\tilde
V(x)\right]x^{\ell+1}\,\tilde w(x)=\frac{2}{\pi}\int\limits_0^\infty{\rm
d}y\,G_\ell(x,y)\,y^{\ell+1}\left[1-\frac{{\rm d}^2}{{\rm
d}y^2}-\frac{2\,(\ell+1)}{y}\,\frac{{\rm d}}{{\rm d}y}\right]\tilde w(y)\
,\label{Eq:IDE}\end{equation}where the kernel $G_\ell(x,y)$ is defined
by$$G_\ell(x,y):=2^\ell\,z^{\ell+1}\left(\frac{1}{z}\,
\frac{\partial}{\partial z}\right)^\ell\frac{1}{z}
\left[(s-z)^{\ell/2}\,K_\ell\left(\sqrt{s-z}\right)
-(s+z)^{\ell/2}\,K_\ell\left(\sqrt{s+z}\right)\right],$$with the
abbreviations $s\equiv x^2+y^2$ and $z\equiv 2\,x\,y.$ Here, $K_\ell$ is the
modified Bessel function~of the second kind of order $\ell$
\cite{Abramowitz}.

\section{Conversion into an Equivalent Matrix Eigenvalue
Problem}\label{Sec:CEMEP}In order to rephrase the eigenvalue equation
(\ref{Eq:IDE}) in the form of an (easier-to-handle) algebraic problem, we
expand every function on the positive real line $R^+$ we encounter into a
complete orthonormal system $\{f_n(x),\ n=0,1,2,\dots\}$ of basis functions
for the Hilbert space $L_2(R^+)$. The solutions $\tilde w(x)$ of
Eq.~(\ref{Eq:IDE}), in particular, are then obtained in the form$$\tilde
w(x)=\sum_{n=0}^Nc_n\,f_n(x)$$ with some set of real coefficients $c_n.$ This
treatment would be exact for $N=\infty$ and represents an approximation for
$N<\infty$ of, however, increasing accuracy with increasing matrix size~$N$.
By application of this procedure, we are able to recast the spinless Salpeter
equation (\ref{Eq:SSE})~into the form of a matrix eigenvalue equation for the
vector $c\equiv\{c_n\}$ of the expansion coefficients:\begin{equation}\tilde
E\,c=\left(P^{(\ell)}\right)^{-1}
\left[\left(T^{(\ell)}\right)^{\rm T}+V^{(\ell)}\right]c\
.\label{Eq:MEE}\end{equation}Here, the elements of the ``power matrices''
$P^{(\ell)}$ and ``potential matrices'' $V^{(\ell)}$ are
defined~by\begin{eqnarray*} P^{(\ell)}_{nm}&:=&\int\limits_0^\infty{\rm
d}x\,x^{\ell+1}\,f_n(x)\,f_m(x)= P^{(\ell)}_{mn}\ ,\quad m,n=0,1,\dots,N\
,\\[1ex] V^{(\ell)}_{nm}&:=&\int\limits_0^\infty{\rm d}x\,x^{\ell+1}\,\tilde
V(x)\,f_n(x)\,f_m(x)=V^{(\ell)}_{mn}\ ,\quad m,n=0,1,\dots,N\
.\end{eqnarray*}The matrices $T^{(\ell)}$ represent the action of the kinetic
term on the vector of basis functions~$f_n.$ The main advantage of our
procedure is the fact that the ``kinetic-energy matrices'' $T^{(\ell)}$ have
to be calculated only once (for a chosen matrix size $N$)! The elements
$T^{(\ell)}_{nm}$ of these matrices, however, have to be calculated
separately for every value of the orbital angular momentum~$\ell.$ The
solution of the matrix equation (\ref{Eq:MEE}) then gives the energy
eigenvalues $E_k,$ $k=0,1,\dots,N,$ as well as the reduced radial wave
functions\begin{equation}w_k(r)=m^{\ell+1}\sum_{n=0}^N\,c_n^{(k)}\,f_n(m\,r)\
,\quad k=0,1,\dots,N\ .\label{Eq:RRWF}\end{equation}

Our choice of the basis functions $f_n(x)$ is primarily dictated by our
demand to allow~for~an analytical treatment of the spinless Salpeter equation
(\ref{Eq:SSE}) to the utmost (reasonable)
extent:$$f_n(x):=\sqrt{2}\,\exp(-x)\,L_n(2\,x)\ ,\quad n=0,1,2,\dots\
,$$where $L_n(x)$ are the Laguerre polynomials \cite{Abramowitz}
$$L_n(x)=\sum_{t=0}^n\,(-1)^t\left(\begin{array}{c}n\\t\end{array}\right)
\frac{x^t}{t!}\ .$$Trivially, these basis functions satisfy the
orthonormalization condition$$\int\limits_0^\infty{\rm
d}x\,f_n(x)\,f_m(x)=\delta_{nm}\ .$$For this choice of basis functions it is
straightforward to write down the explicit expression~for the matrix elements
$P^{(\ell)}_{nm}$:
$$P^{(\ell)}_{nm}=\frac{1}{2^{\ell+1}}\,\sum_{p=0}^n\,\sum_{q=0}^m\,(-1)^{p+q}
\left(\begin{array}{c}n\\p\end{array}\right)
\left(\begin{array}{c}m\\q\end{array}\right)
\frac{(p+q+\ell+1)!}{p!\,q!}\ .$$This easy availability of the explicit
expressions for the power matrices $P^{(\ell)}$ strongly suggests to
consider, as a special case, a class of interaction potentials of
(generalized) power-law~form:\begin{equation}V(r)=\sum_{n\in Z}\,a_n\,r^n\
,\label{Eq:GPLP}\end{equation}where the two sets of (otherwise arbitrary)
integers $n$ and real constants $a_n$ (playing the~r\^ole of coupling
strengths) are only constrained~by the necessary boundedness from below of
the Hamiltonian $H$ defined in Eq.~(\ref{Eq:SRH}): $n\ge-1$ if $a_n<0.$ For
this class of interaction potentials, the potential matrices $V^{(\ell)}$
simply become linear combinations of the power matrices
$P^{(\ell)}$:$$V^{(\ell)}=\sum_{n\in Z}\,\frac{a_n}{m^{n+1}}\,P^{(\ell+n)}\
.$$It goes without saying that the procedure described here works in the same
way also, at~least, for all power-law potentials involving additional
exponential (damping) factors, that is, for~all interaction potentials of the
form $$V(r)=\sum_{n\in Z}\,a_n\,r^n\,\exp(-b_n\,r)\ ,\quad b_n\ge0\ .$$

\section{The Kinetic-Energy Matrix}\label{Sec:KEM}Clearly, the main task in
our game is the calculation of the kinetic-energy matrix $T^{(\ell)}.$ This
enterprise is somewhat involved but, as already mentioned, has to be
undertaken only once. In general, the elements of the kinetic-energy matrix
$T^{(\ell)}$ are given by some expression of~the form \cite{Lucha92NAM}
$$T^{(\ell)}_{nm}=\frac{2}{\pi}\,\sum_{k=0}^n\,D(\ell;n,k)\,c(\ell;k,m)\
,\quad m,n=0,1,\dots,N\ .$$Here, the factors $D(\ell;n,k)$ represent the
action of the differential operator on the right-hand side of the
integro-differential equation (\ref{Eq:IDE}) on the chosen basis functions
$f_n(x).$ For~our~choice of basis functions, these factors read explicitly
\cite{Lucha92NAM}$$D(\ell;n,k)\equiv
2\,\sqrt{2}\,\frac{(-1)^k}{k!}\,2^k\left[(k+\ell+1)
\left(\begin{array}{c}{n}\\{k}\end{array}\right)
+(k+2\,\ell+2)\left(\begin{array}{c}{n}\\{k+1}\end{array}\right)\right].$$The
factors $c(\ell;k,m)$ are the expansion coefficients, in terms of the basis
functions $f_n(x),$~of~a particular integral over the kernel $G_\ell(x,y)$
\cite{Lucha92NAM}:$$c(\ell;k,m)=\int\limits_0^\infty{\rm d}x\,f_m(x)\,
\int\limits_0^\infty{\rm d}y\,G_\ell(x,y)\,y^{\ell+k}\,\exp(-y)\ .$$For
increasing orbital angular momentum $\ell$, the kernels $G_\ell(x,y),$
$\ell=0,1,2,\dots,$ become~very rapidly rather complicated expressions. This
circumstance forces us to calculate these factors $c(\ell;k,m)$ separately
for every single value of the orbital angular momentum $\ell=0,1,2,\dots$~we
are interested in. Here, we confine ourselves to the discussion of the cases
$\ell=0$ and $\ell=1.$ We find, for states with orbital angular momentum
$\ell=0$ (the so-called ``S waves''), \cite{Lucha92NAM}\begin{eqnarray*}
c(0;k,m)&=&2\,\sqrt{2}\,k!\,\sum_{p=0}^{[k/2]}\,\sum_{q=0}^p\,\sum_{r=0}^m\,
(-1)^{q+r}\left(\begin{array}{c}k+1\\2\,p+1\end{array}\right)
\left(\begin{array}{c}p\\q\end{array}\right)
\left(\begin{array}{c}m\\r\end{array}\right)\\[1ex]&\times&
\frac{2^r\,(r+1)\,(2\,k-2\,q+r+1)!}{(2\,k-2\,q+1)!!\,(2\,k-2\,q+2\,r+3)!!}\
,\end{eqnarray*}and, for states with orbital angular momentum $\ell=1$ (the
so-called ``P waves''), \cite{RupprechtPhD}
\begin{eqnarray*}c(1;k,m)&=&2\,\sqrt{2}\,k!\,
\sum_{p=0}^{[k/2]+1}\,\sum_{q=0}^{p+1}\,\sum_{r=0}^m\,(-1)^{q+r}\left[
\left(\begin{array}{c}k+2\\2\,p\end{array}\right)
\left(\begin{array}{c}p\\q\end{array}\right)
\frac{k+1}{(2\,k-2\,q+3)!!}\right.\\[1ex]&+&\left.\left(
\begin{array}{c}k+1\\2\,p+1\end{array}\right)
\left(\begin{array}{c}p\\q-1\end{array}\right)
\frac{1}{(2\,k-2\,q+5)!!}\right]\left(\begin{array}{c}m\\r\end{array}\right)
\frac{2^r\,(2\,k-2\,q+r+4)!}{(2\,k-2\,q+2\,r+5)!!}\ ,\end{eqnarray*}
where$$\left[\frac{k}{2}\right]\equiv\left\{\begin{array}{ll}
\displaystyle\frac{k}{2}&\quad\mbox{for $k$ even}\ ,\\[2ex]
\displaystyle\frac{k-1}{2}&\quad\mbox{for $k$ odd}\
,\end{array}\right.$$and$$ (2\,n+1)!!\equiv 1\times
3\times\cdots\times(2\,n-1)\times(2\,n+1)\ ,\quad n=0,1,2,\dots\ .$$The
calculation of the factors $c(\ell;k,m)$ for $\ell>1$ is straightforward but
somewhat involved because of the increasing complexity of the kernels
$G_\ell(x,y).$

\section{The Mathematica 4.0 Notebook {\tt Salpeter}.}\label{Sec:MNBS}The
``semianalytical matrix method'' developed here for the solution of the
spinless Salpeter equation (\ref{Eq:SSE}) has been implemented in a
Mathematica 4.0 routine called {\tt Salpeter.nb} (which may be obtained by
contacting the authors). The routine {\tt Salpeter.nb} requires as input,~for
two particles of mass $m$ experiencing some interaction potential $V(r)$ and
for a chosen matrix size $d\equiv N+1,$ only this potential $V(r)$ expressed
in terms of the single terms $a_n\,r^n$ in Eq.~(\ref{Eq:GPLP}), these terms
written in the form {\tt vpot[asubn,n,m,d]}. For instance, for the ``funnel''
potential$$V(r)=-\frac{a_{-1}}{r}+a_1\,r\ ,$$type\\[1ex]\noindent{\tt
v[m\_,d\_]:=vpot[-asubminus1,-1,m,d]+vpot[asub1,1,m,d]}.\\[1ex]\noindent Upon
entering the command {\tt etot[ell,m,d]}, this routine then computes, for
orbital angular momentum $\ell,$ particle mass $m,$ and matrix size $d\equiv
N+1,$ the bound-state energy eigenvalues $E_k,$ $k=0,1,\dots,N,$ together
with the eigenvectors $c^{(k)},$ $k=0,1,\dots,N,$ from Eq.~(\ref{Eq:MEE})
and~the corresponding radial eigenfunctions $R_k(r)\equiv r^\ell\,w_k(r),$
$k=0,1,\dots,N,$ according to Eq.~(\ref{Eq:RRWF}).

We illustrate the power of the ``semianalytical matrix method'' by applying
it to the~case of a harmonic-oscillator potential $V(r)=a\,r^2,$ $a>0,$ for
the following reason: In momentum space, the operator $r^2$ is represented by
the Laplacian w.~r.~t. the momentum~${\bf p}$, $r^2\rightarrow-\Delta_{\bf
p}$, while the kinetic energy $T$, nonlocal in configuration space, is
represented by a multiplication operator. Only for a harmonic-oscillator
potential the momentum-space representation of~the semirelativistic
Hamiltonian $H$ is therefore of the form of a nonrelativistic Hamiltonian
with a square-root interaction potential:$$H=-a\,\Delta_{\bf p}+2\,\sqrt{{\bf
p}^2+m^2}\ .$$The resulting (nonrelativistic) Schr\"odinger equation is then
solved with a standard numerical procedure designed exactly for this purpose
\cite{Lucha99:SEM}. (For details, in particular, for exact analytical upper
and lower bounds on the energy levels, see, for instance,
Refs.~\cite{Lucha99:QA}.)

\small\begin{table}[h]\caption[]{\small Eigenvalues (in units of GeV) of the
semirelativistic Hamiltonian $H=2\,\sqrt{{\bf p}^2+m^2}+V(r)$ with a
harmonic-oscillator potential $V(r)=a\,r^2$, for states of principal quantum
number $n=1,2,3,4$ and orbital angular momentum $\ell=0$ or $\ell=1$ (called
$n$S or $n$P in the usual spectroscopic notation) obtained by the
``semianalytical matrix method'' for increasing matrix sizes $d=1,3,10,25,$
compared with the corresponding ``exact'' eigenvalues obtained from the
momentum-space representation of~the spinless Salpeter equation, which, for
the special case of a harmonic-oscillator potential, resembles a
nonrelativistic Schr\"odinger equation (as discussed in the text). The chosen
physical parameter values are $m=1\;\mbox{GeV}$ for the particle mass and
$a=0.5\;\mbox{GeV}^3$ for the harmonic-oscillator coupling
strength.}\label{Tab:HOP}\small
\begin{center}\begin{tabular}{rrrrrrrrr}\hline\hline&&&&&&&&\\[-1.5ex]
\multicolumn{1}{c}{$d=N+1$}&\multicolumn{4}{c}{$\ell=0$}
&\multicolumn{4}{c}{$\ell=1$}\\[1ex]
\cline{2-9}\\[-1.5ex]&\multicolumn{1}{c}{1S}&\multicolumn{1}{c}{2S}&
\multicolumn{1}{c}{3S}&\multicolumn{1}{c}{4S}
&\multicolumn{1}{c}{1P}&\multicolumn{1}{c}{2P}&
\multicolumn{1}{c}{3P}&\multicolumn{1}{c}{4P}\\[1ex]\hline\\[-1.5ex]
1&4.14531&\multicolumn{1}{c}{---}&\multicolumn{1}{c}{---}&
\multicolumn{1}{c}{---}&5.12166&\multicolumn{1}{c}{---}&
\multicolumn{1}{c}{---}&\multicolumn{1}{c}{---}\\[1ex]
3&3.91571&7.08622&11.94458&\multicolumn{1}{c}{---}
&5.03946&7.68220&14.80138&\multicolumn{1}{c}{---}\\[1ex]
10&3.82522&5.80930&7.76609&9.90246&4.89944&6.72710&8.54739&11.20168\\[1ex]
25&3.82494&5.79112&7.48323&9.01617&4.90149&6.69298&8.28585&9.74304\\[1ex]
\hline\\[-1.5ex]
exact&3.82493&5.79102&7.48208&9.00749&4.90145&6.69305&8.28464&9.74276\\[1ex]
\hline\hline\end{tabular}\end{center}\end{table}\normalsize

Table~\ref{Tab:HOP} demonstrates the rapid convergence for increasing matrix
size of the lowest-lying energy eigenvalues of the spinless Salpeter equation
(\ref{Eq:SSE}) with a harmonic-oscillator potential.

\end{document}